\documentclass[aps,prx,twocolumn,floatfix,longbibliography,nofootinbib,tbtags]{revtex4-1}
\usepackage[utf8]{inputenc}
\usepackage{natbib}
\usepackage{graphicx}
\usepackage{xcolor}
\usepackage{mathtools}
\usepackage{amsmath}
\usepackage[colorlinks, linkcolor=blue]{hyperref}
\hypersetup{colorlinks,allcolors=black}
\usepackage{amssymb}
\usepackage{gensymb}
\usepackage{float}

\usepackage{tabularx,graphicx}
\usepackage{epstopdf}
\usepackage{latexsym}
\usepackage{color, colortbl}
\usepackage{psfrag}
\usepackage{bbm}
\usepackage{bm}
\usepackage{titlesec}
\usepackage{dsfont}
\usepackage{feynmp}
\usepackage{slashed}
\usepackage{multirow}
\usepackage[normalem]{ulem}
\renewcommand{\vec}[1]{\boldsymbol{#1}}
\def \nn {\nonumber}

\def \k {{\vec k}}

\def \ve {\varepsilon}
\def \r {{\vec r}}

\def \v {{\bf v}}

\def \q {{\vec q}}

\def \l {\ell}
\def \ve {\varepsilon}

\def \l {\ell}

\def \n{{\mathbf{n}}}

\def \beq {\begin{eqnarray}}
\def \eeq {\end{eqnarray}}
\def \tn {\textnormal}

\def \m {{\mathcal}}

\def \la{\langle}
\def \ra{\rangle}

\def \hk {H_{\tn{kin}}}
\def \hi {H_{\tn{int}}}
\def \kxx {K_{\tn{xx}}^{\tn{eff}}}
\def \kxxnaive {K_{\tn{xx}}^{\tn{naive}}}
\def \J {J^{\tn{eff}}}
\def \cxx {\chi_{\tn{xx}}^{\tn{eff}}}

\def \heff {\mathcal{H}_{\tn{eff}}}

\begin{document}
\title{Diamagnetic response and phase stiffness for interacting isolated narrow bands}
\author{Dan Mao}
\author{Debanjan Chowdhury}
\affiliation{Department of Physics, Cornell University, Ithaca, New York 14853, USA.}

\begin{abstract}
Superconductivity is a macroscopic manifestation of a quantum phenomenon where pairs of electrons delocalize and develop phase coherence over a long distance. A long-standing quest in both theory and experiment has been to address the underlying microscopic mechanisms that fundamentally limit the superconducting transition temperature, $T_c$. A platform which serves as an ideal playground for realizing ``high'' temperature superconductors are materials where the electrons' kinetic energy is completely quenched  and interactions provide the only energy-scale in the problem for $T_c$. However, when the non-interacting bandwidth for a set of isolated bands is small compared to the scale of the interactions, the problem is inherently non-perturbative and requires going beyond the traditional mean-field theory of superconductivity. In two spatial dimensions, $T_c$ is controlled by the superconducting phase stiffness. Here we present a general theoretical framework for computing the electromagnetic response for generic model Hamiltonians, which controls the maximum possible superconducting phase stiffness and thereby $T_c$, without resorting to any mean-field approximation. Importantly, our explicit computations demonstrate that the contribution to the phase stiffness arises from (i) ``integrating-out'' the remote bands that couple to the microscopic current operator, and (ii) the density-density interactions projected on to the isolated narrow bands. 
Our framework can be used to obtain an 
upper bound on the phase stiffness, and relatedly the superconducting transition temperature, for a range of physically inspired models involving both topological and non-topological narrow-bands with arbitrary density-density interactions. We discuss a number of salient aspects of this formalism by applying it to a specific model of interacting flat-bands and compare against the known $T_c$ from independent numerically exact computations. 
\end{abstract}

\maketitle
\begin{figure}[!h]
\centering
\includegraphics[width=1.0\linewidth]{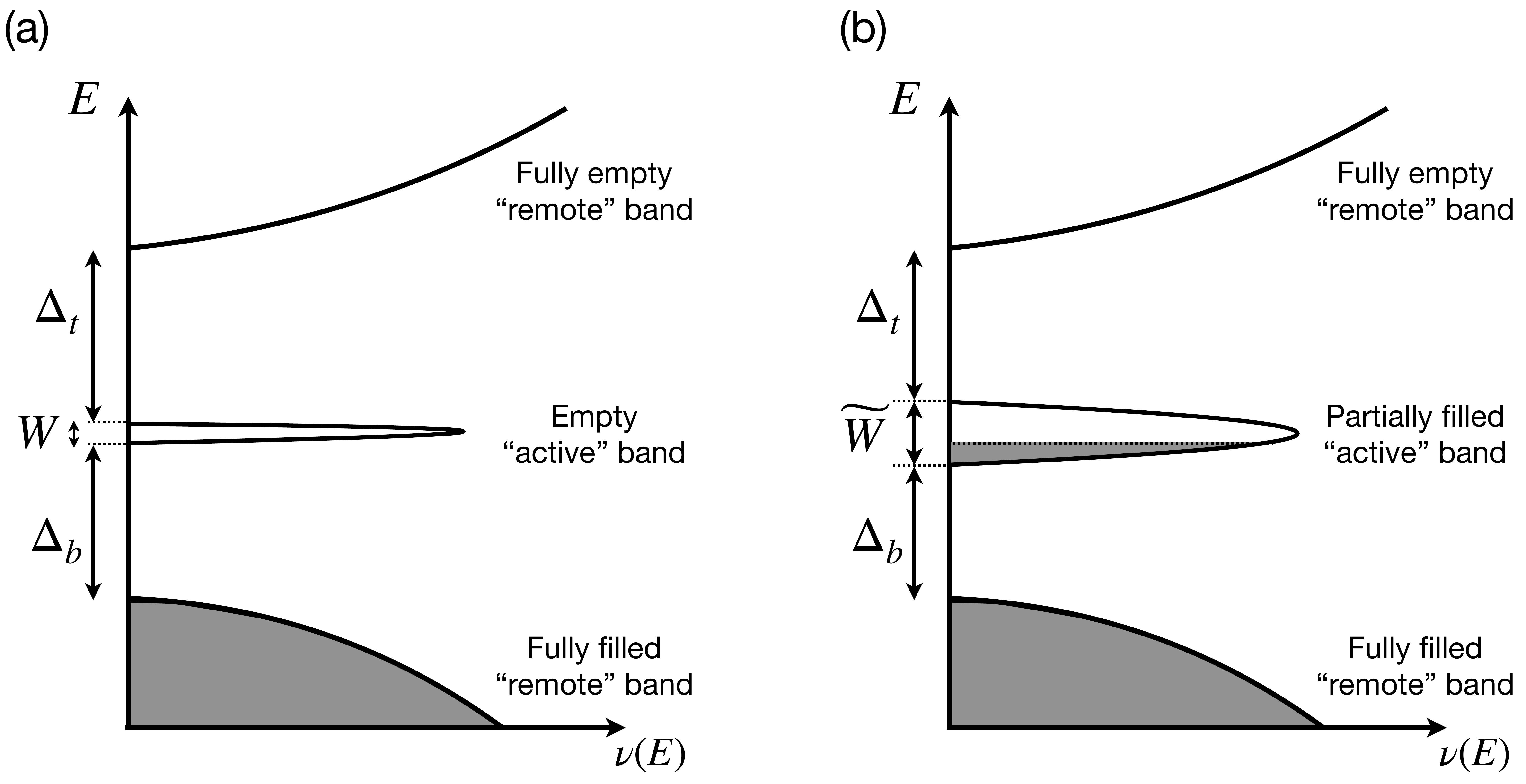}
\caption{A schematic of the energy-bands in the (a) non-interacting, and (b) interacting limits, respectively. In (a), we denote the bandwidth associated with the isolated narrow bands as $W$, which is well separated from the remote (bottom and top) bands by gaps, $\Delta_b,~\Delta_t$, respectively. In (b) the interaction-induced renormalization of the isolated bands leads to a modified bandwidth, $\widetilde{W}$. We assume a hierarchy of energy scales, $W\lesssim V(\sim\widetilde{W})\ll \Delta_{b,t}$. This manuscript derives the effective diamagnetic response, $\kxx$, and the integrated optical spectral weight, $\int_0^{\Lambda} d\omega~\tn{Re}[\sigma_{\tn{xx}}^{\tn{eff}}(\omega)]$ (where $\widetilde{W}<\Lambda<\Delta_t$), for the isolated bands with projected density-density interactions in the limit $\Delta\rightarrow\infty$.}
\label{bands}
\end{figure}

What is the highest attainable superconducting temperature $T_c$ in a given physical system? One suggested theoretical route to enhance $T_c$ has been to focus on the problem of ``flat-band'' superconductivity \cite{Shaginyan1990,Kopnin2011,Volovik2013}--- a strong to intermediate-coupling regime, where Bardeen-Cooper-Schrieffer (BCS) mean-field theory does not {\it a priori} apply. In particular, attempting to enhance the gap-scale in the strong-coupling regime often comes at the cost of degrading the phase-coherence scale \cite{MR10}. However, the discovery of superconductivity across a number of two-dimensional moir\'e materials in the last few years  \cite{Cao2018,AY19,Efetov19,SNP20,PK21,park2021tunable} has refocused our attention on this question. 
Across most of these platforms, the low-energy physics is associated with a set of partially filled ``active'' bands, which in the non-interacting limit have a narrow bandwidth, $W$. These active bands are well-separated from the ``remote'' bands by a charge-gap, $\Delta$, and the characteristic electronic interaction scale, $V$, is believed to be $W\lesssim V\ll\Delta$; see Fig.~\ref{bands}. While the detailed microscopic mechanism for the origin of superconductivity across these materials remains unclear, there are basic conceptual and model-independent questions that have not been understood going beyond BCS theory. In two-dimensions and in the limit of strong interactions, $T_c$ is limited by the superconducting phase stiffness \cite{NK77,emery1995importance}, which is determined by the transverse electromagnetic response of the electronic system \cite{scalapino1993insulator}. One of the goals of this paper is to lay down the theoretical foundation for 
analyzing the low-energy electromagnetic response for a narrow-bandwidth system in the regime of strong interactions and computing the largest possible superconducting phase stiffness when interactions are projected to the isolated flat-bands. An additional level of complexity arises from the non-trivial momentum dependence of the Bloch wavefunctions associated with the active bands, which affect the nature of interactions projected to these bands, as well as the effective current correlations obtained upon integrating out the higher-energy degrees of freedom. Characterizing the superconducting instabilities in this non-perturbative regime across their respective phase-diagrams, and their detailed dependence on the various microscopic parameters, requires the development of new analytical and numerical methods.  

Since a reliable microscopic theory of superconductivity for interacting flat-bands is currently unavailable, an interesting point of view has been to formulate fundamental ``bounds'' on the superconducting phase-stiffness, $D_s(T)$, at a temperature, $T$. In two-dimensions, the superconducting $T_c$ and $D_s(T)$ are related to each other, $T_c = \pi D_s(T_c^-)/2$ \cite{NK77}. Within BCS mean-field theory ({\it a priori} unjustified) and in the superconducting ground state (i.e. ignoring competing instabilities), $D_s(0)$ is bounded from below by a geometric invariant \cite{Torma15,Torma19,Bernevig19,Rossi19,Bernevig_review}, namely the Fubini-Study metric, associated with the Bloch wavefunctions \cite{VanderbiltRMP}. Within a set of restrictive conditions, the BCS wavefunction can be shown to be an exact ground-state \cite{Huber16,JHA22}, but $T_c=0$ as a result of an emergent SU(2) symmetry. However, numerically exact solution to the many-body problem in the presence of infinitesimal changes to the Hamiltonian are known to induce non-superconducting competing orders \cite{hofmann2020superconductivity,hofmann2022superconductivity}, where $D_s(0)=0$. On the other hand, given the entire bounded electronic spectrum (i.e. including all the bands) for any microscopic Hamiltonian with purely density-density interactions, the optical spectral weight integrated upto the full bandwidth serves as a rigorous upper bound on $D_s$ \cite{hazra2019bounds}. For most electronic solids, this will typically be $D_s\lesssim O(\tn{eV})$. This bound is useful for Galilean-invariant systems (especially at low-densities), where $T_c$ is bounded by the Fermi-energy, $E_F$ (upto a numerical prefactor). More generally, there are no analogous bounds on $T_c/E_F$, or on $T_c/D_s(0)$ in non-Galilean-invariant systems \cite{heuristicbound}, which necessitates the development of a more widely applicable theoretical framework for addressing upper-bounds on $T_c$. 

Deriving a rigorous ``low-energy'' upper bound on $T_c$ in the limit $\Delta\rightarrow\infty$, which is controlled by an interacting effective Hamiltonian, $\heff$, acting only on the active bands remains an outstanding challenge. This manuscript will be concerned with formulating the theoretical framework that is necessary to derive such an upper bound. This program depends on the strength and form (i.e. spatial profile) of the interactions. As a result, it will depend ultimately on the nature of the various multi-particle correlation functions within the low-energy manifold of states which is acted upon by $\heff$.  
A number of recent works have directly studied the many-body problem in sign-problem-free models using numerically exact quantum Monte-Carlo (QMC) computations \cite{hofmann2020superconductivity,Huber21,Bernevig21,Zhang2021,hofmann2022superconductivity}. As a concrete example, we apply our framework to derive a theoretical upper bound on $D_s(0)$ for a model of interacting flat-bands below, where the exact result is known independently from QMC.

{\bf Model.-} In general, $D_s$ can be computed as the {\it transverse} electromagnetic response \cite{scalapino1993insulator,resta2018drude} for an effective Hamiltonian, $\heff[A]$, in the presence of a probe gauge-field, $A$,

\begin{subequations}
 \beq
        \frac{D_s}{\pi e^2} &=& \bigg[\langle \kxx \rangle - \cxx(\omega=q_x=0,q_y \rightarrow 0)\bigg]_{A\rightarrow0}, \label{Ds}\\
       \kxx &=& \frac12 \frac{\delta^2 \heff[A]}{\delta A_\tn{x} \delta A_\tn{x}},~~
        \cxx(\q) = \langle \J_\tn{x}(\q)~\J_\tn{x}(-\q)\rangle, \nn\label{susc}\\
\label{eq:D}
\eeq
\end{subequations}
where $\kxx$ is the {\it effective} diamagnetic contribution and $\cxx$ is the {\it effective} current susceptibility, with $\J_\tn{x} = -\delta \heff[A]/\delta A_{\tn{x}}$. The expectation values are evaluated with respect to the many-body state at inverse temperature $T^{-1}=\beta$ as $\langle ...\rangle = \tn{Tr}[e^{-\beta\heff}...]/\tn{Tr}[e^{-\beta\heff}]$. One of the important contributions of our work will be explicit low-energy formulas for these effective operators and response functions, where we will highlight their key differences from the naive (microscopic) definitions for the same. 

Given a UV Hamiltonian, $H$,  which is distinct from $\heff$, there are a number of subtleties associated with computing these effective susceptibilities. To be concrete, we start with a generic translationally invariant UV Hamiltonian,
\begin{subequations}
\beq
H &=& \hk + \hi, \label{ham1}\\
\hk &=& \sum_{\substack{\r,\r'\\ \alpha,\alpha'}} t_{\alpha\alpha'}(\r-\r') c^\dagger_{\r\alpha} c^{\phantom\dagger}_{\r'\alpha'} - \mu N,\label{ham2}\\
\hi &=& \sum_{\r,\r'} V(\r-\r')~ n_\r n_{\r'},
\eeq
\end{subequations}
where $c_{\r\alpha},~c^\dagger_{\r\alpha}$ denote microscopic electronic operators at site $\r$ with a collective orbital and spin index $\alpha$, satisfying the standard anti-commutation algebra. In $\hk$, the matrix $t_{\alpha\alpha'}(\r-\r')$ includes a set of arbitrary inter-site/orbital hoppings and $\mu$ is an external chemical potential that couples to $N = \sum_{\r}n_\r = \sum_{\r,\alpha} c^\dagger_{\r\alpha} c^{\phantom\dagger}_{\r\alpha}$. For simplicity, in the remainder of this work we only consider density-density interactions, $V(\r-\r')$, at the UV scale (e.g. we ignore electron-phonon interactions, ``correlated-hopping'', pair-hopping interactions etc.). An upper bound on the phase stiffness for $H$ can be evaluated --- the probe gauge field {\it only} couples to $\hk$, leading to an upper bound for $D_s$ expressed in terms of the full optical spectral weight \cite{hazra2019bounds}.  

By transforming from the orbital to band basis, we obtain
$\hk= \sum_{\k,m} (\ve_{\k m} - \mu) c^\dagger_{\k m} c^{\phantom\dagger}_{\k m}$,
where $m$ represents (possibly degenerate) bands with Bloch functions, $|u_{\k,m}\ra$. Then, let us recall that the microscopic current operator and the diamagnetic response can be obtained as \cite{scalapino1993insulator},

\beq
        J_\mu(\q) &=& \sum_{\k,m,m'} c_{\k+\frac{\q}{2}m}^\dag c^{\phantom\dagger}_{\k-\frac{\q}{2}m'}
    \langle u_{\k+\frac{\q}{2},m} | \partial_{k_\mu} \hat{h}_\k|u_{\k-\frac{\q}{2},m'}\rangle, \nn\\
        K_{\mu\nu} &=& \sum_{\k,m,m'} c_{\k m}^\dag c^{\phantom\dagger}_{\k m'} \langle u_{\k,m} | \partial_{k_\mu}\partial_{k_\nu} \hat{h}_\k|u_{\k,m'}\rangle,
        \label{eq:JandK}
\eeq
where $\hat{h}_\k = \sum_n (|u_{\k,n}\rangle \ve_{\k n} \langle u_{\k n}|)$ is the matrix of $(\hk+\mu N)$.

For later convenience, note that the diamagnetic term and the current operator (in the small $\q$ limit) can also be written in the following way,
\begin{subequations}
\beq
J_\mu(q_\mu \rightarrow 0) &=&  -i \left[\hat{X}_\mu, H\right],\\
K_{\mu\nu} &=& -\left[\hat{X}_\mu, \left[\hat{X}_\nu, H\right]\right],
\eeq
\end{subequations}
where $\hat{X}_\mu \equiv \sum_i x_i^\mu c_i^\dag c_i$ is the many-body position operator.






For the remainder of our discussion, we will only be interested in the effective low-energy contribution to $D_s$ from a finite subset, $\{m\in\tn{active~bands}\}$, that are separated from the remote bands by a large gap (Fig.~\ref{bands}). There are two conceptually equivalent approaches for addressing this problem: (1) In the limit where $\Delta$ is finite but large, the microscopic current operator couples together the active and remote bands; the effects of the remote bands have to be ``integrated-out'' via a Schrieffer-Wolff (SW) transformation \cite{SW_rev}. (2) The model has a global conserved $U(1)$ density associated with the total particle number. However, in the limit where $\Delta\rightarrow\infty$, there is an {\it emergent} conservation law associated with the electronic density {\it restricted} to the active bands, and a corresponding current operator. The electromagnetic response can then be obtained by carrying out an appropriately defined gauge-transformation in terms of ``projected'' coordinates, that does not take us out of the low-energy Hilbert space. One caveat is that by taking the second approach, one has to first determine how the physical electromagnetic response is related to the response to the emergent gauge transformation, which is {\it a priori} unknown. Within both approaches, it is worth noting that the projected interaction can generate a finite bandwidth, which in turn contributes to $D_s$.

\begin{figure}
    \centering  \includegraphics[width=0.50\textwidth]{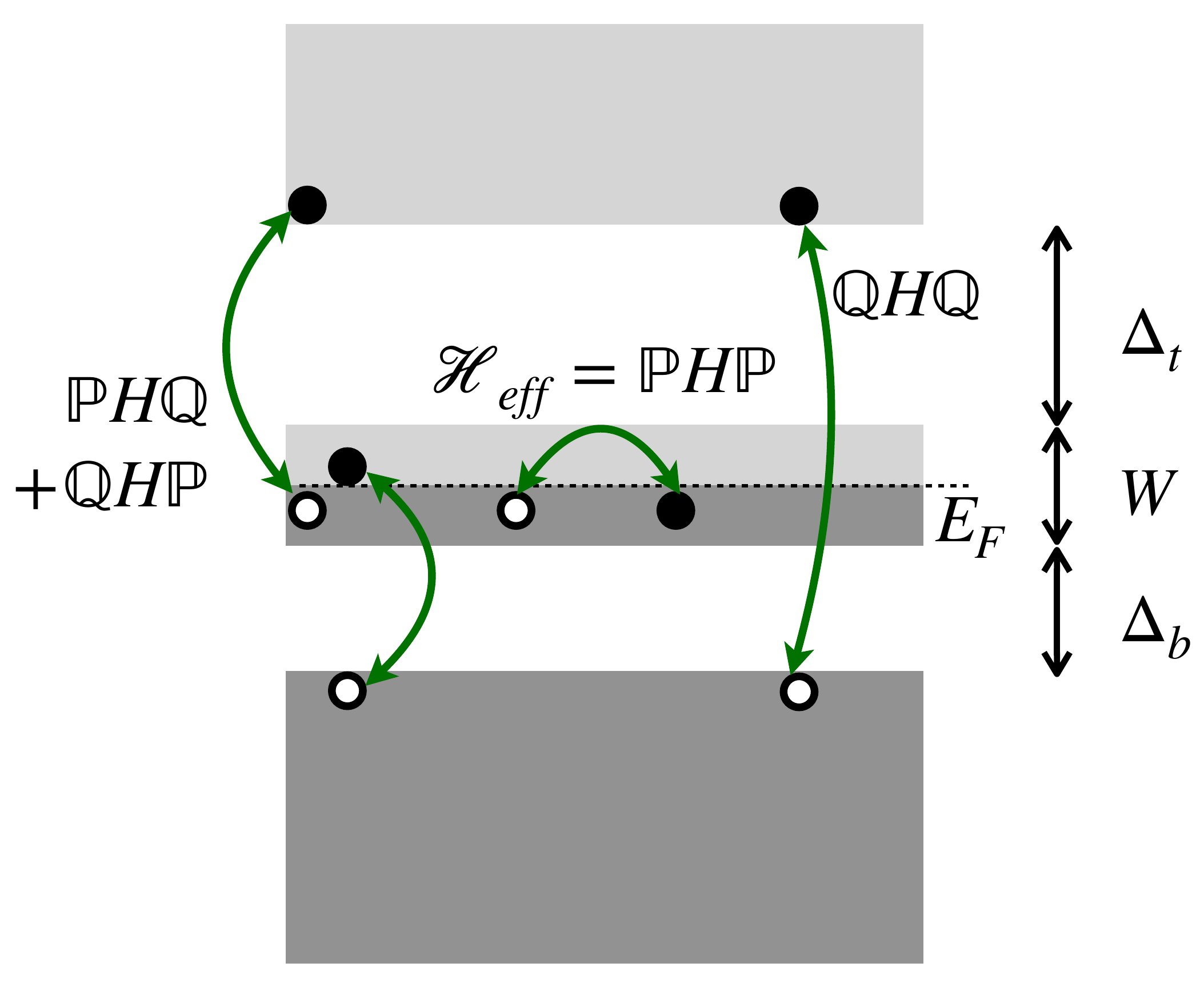}
    \caption{
    A schematic depiction of the projections and Schrieffer-Wolff transformations. $\heff$ is the effective Hamiltonian which acts only within the low-energy sub-Hilbert space obtained by the action of the many-body projector, $\mathbb{P}$. The term $\mathbb{Q} H \mathbb{Q}$ acts purely outside the sub-Hilbert space spanned by $\mathbb{P}$. The off-diagonal term, $H_o = \mathbb{P} H \mathbb{Q} + \mathbb{Q} H \mathbb{P} $, mixes the low-energy states with the high-energy states. }
    \label{fig:proj}
\end{figure}

We will begin by adopting the first approach, where we can analyze the effects of a finite $\Delta$ on the physics explicitly. Let us resolve the Hamiltonian into its ``diagonal'' and ``off-diagonal'' pieces, respectively,
\begin{subequations}
\beq
H &=& H_d + H_o, \\
H_d &=& \mathbb{P} H \mathbb{P} + \mathbb{Q} H \mathbb{Q},\\
H_o &=& \mathbb{P} H \mathbb{Q} + \mathbb{Q} H \mathbb{P}, 
\eeq
\end{subequations}
where $\mathbb{Q}=\mathbb{I}-\mathbb{P}$.  Here, $\mathbb{P}$ is the projection operator to the sub-Hilbert space, $\mathbb{H}$, spanned by the many-body states with partially occupied active bands and fully occupied (empty) lower energy (higher energy) remote bands;  see Fig.~\ref{fig:proj} for a schematic depiction of the action of different projections. The basis of $\mathbb{H}$ with $n$ occupied states in the active bands can be constructed from the ``vacuum'' state, $|\psi_0\rangle$, corresponding to fully filled remote bands as,
\beq
|\psi^{(n)}\rangle = \prod_{p=1}^{n} c_{\k_p m_p}^\dag |\psi_0\rangle.
\eeq 
Here the set of $m_p$ belongs to the set of active bands and the state $|\psi^{(n)}\rangle$ has total momentum, $\sum_p \k_p$. Since the total number of particles is conserved, any many-body state that does not belong in this low-energy manifold necessarily has holes (electrons) in the lower (upper) bands, and is therefore well separated in energy. Moreover, later when we consider the limit of $\Delta\rightarrow\infty$, there is an emergent conservation law associated with the density of electrons in the active bands. Finally, we note that the low-energy Hamiltonian of interest to us is $\heff \equiv \mathbb{P} H\mathbb{P} = \mathbb{P} H_d\mathbb{P}$.

{\bf Results.-} For a small $A$, we expand $H[A] = H[0] + J_\mu A_\mu + \frac12 K_{\mu\nu} A_\mu A_\nu + ... \equiv H_d[A] + H_o[A]$, where $J_\mu$ and $K_{\mu\nu}$ are defined in Eq.~\ref{eq:JandK}. Clearly, $H_o[A]$ will introduce mixing between the active and remote bands. To leading order in $H_o[A]$, we can use the SW transformation to obtain a unitarily equivalent Hamiltonian, $\widetilde{H}[A] = e^{ T[A]} H[A] e^{-T[A]}$, which will have no matrix elements between the active and remote bands \cite{SW_rev}. The relevant matrix elements, $\langle m | T[A] |n \rangle = \langle m |H_o[A] |n\rangle/(E_m - E_n)$, where $m,n$ correspond to energy levels of $H_d[A]$ with $|E_m - E_n| \geq \Delta$. This leads us to the $\heff[A]\equiv \mathbb{P}\widetilde{H}[A]\mathbb{P}$, that was introduced in Eq.~\ref{Ds}:
\beq
\heff[A] &=&  \mathbb{P}H_d[A] \mathbb{P}    \label{eq:H_eff_SW}\\
& +&\frac{1}{2}  \sum_{\substack{m,n\in\mathbb{H},\\ \l\notin \mathbb{H}}} \bigg[\langle m |H_o[A] |\l\rangle \langle \l| H_o[A] |n\rangle \times\nn\\ 
&&~~~~~~~~~~~\left(\frac{1}{E_m -E_\l} - \frac{1}{E_\l - E_n}\right)\bigg]  + ...,\nn
\eeq
 where we expand $\widetilde{H}[A]$ using the Baker-Hausdorff-Campbell formula and keep only the leading order term. Higher order terms are supressed by either $1/\Delta$ or of higher order than $A^2$. We can now obtain the effective current, $\J_\tn{x}$, and diamagnetic contribution, $\kxx$, respectively, by expanding Eq.~\ref{eq:H_eff_SW} up to second order in $A$ and calculating the appropriate derivatives as in Eq.\ref{susc}. For $A=0$, the first term in Eq.~\ref{eq:H_eff_SW} ($=\mathbb{P}H\mathbb{P}$) is independent of $\Delta$, while the second term is  $O(V^2/\Delta)$. On the other hand, for $A\neq 0$, it is important to note that both the first and second terms in Eq.~\ref{eq:H_eff_SW} contribute an $O(\Delta)$ correction to the $O(A^2)$ terms. Explicitly, the first term contributes $\mathbb{P} \hat{X} \mathbb{Q} \hk \mathbb{Q} \hat{X} \mathbb{P} \sim O(\Delta)$ and the second term contributes $\mathbb{P} J_x \mathbb{Q}J_x\mathbb{P}/\Delta \sim O(\Delta)$ corrections, respectively. However, these contributions cancel exactly when combined together and to the leading order $\langle \kxx\rangle$ is independent of $\Delta$.  This cancellation is indeed expected, since the result should be finite even in the limit $\Delta\rightarrow\infty$. The detailed derivation and explicit demonstration of the cancellation are discussed in the Supplementary Material \cite{si}.

In simplified form, the effective current operator becomes (after dropping a correction $\sim O(V^2/\Delta)$),
\begin{equation}
    \J_\mu(\q \rightarrow 0) = \mathbb{P} \left(J_\mu (\q\rightarrow 0)  + i \left[\hat{X}_\mu, H_o\right]\right) \mathbb{P},
    \label{eq:J_eff}
\end{equation}
where the first term is just the projected microscopic current, while the second term arises from mixing between active and remote bands \cite{sondhi1992long}. 
Since $\hi$ is a density-density interaction, $[\hat{X}_\mu, \hi] =0$, and the second term in Eq.~\ref{eq:J_eff} can  be rewritten as $-i \left[\mathbb{P}\hat{X}\mathbb{P}, \mathbb{P} \hi \mathbb{P}\right]$, which means that $\J_\mu(\q \rightarrow 0)$ can be expressed in terms of operators acting only within $\mathbb{H}$.  Moreover, given the positive semi-definite property of $\cxx$ \cite{hazra2019bounds}, it is possible to express an upper bound on $D_s$ in terms of the $\Delta-$independent $\langle \kxx\rangle$, which is exactly related to the diamagnetic response for the problem formulated directly in the projected limit (i.e. $\Delta\rightarrow\infty$).

The effective diamagnetic response is given by (upon retaining terms $\sim O(1)$),
\beq
   \langle \kxx \rangle =- \bigg\langle\left[\mathbb{P}\hat{X}\mathbb{P}, \left[\mathbb{P}\hat{X}\mathbb{P}, \mathbb{P} H_d \mathbb{P}\right]\right]\bigg\rangle.
   \label{eq:K_xx_gauge}
\eeq
The above result is already suggestive of an important subtlety, namely that it is only the projected degrees of freedom that enter $\kxx$. This can be seen in a more transparent fashion by re-expressing it as,
\begin{subequations}
\beq
 &&\langle\kxx\rangle = \lim_{\alpha \rightarrow 0} \partial_\alpha^2 \langle e^{i \alpha \mathbb{P} \hat{X} \mathbb{P}} H_d e^{-i \alpha \mathbb{P} \hat{X}\mathbb{P}} \rangle \label{projgauge}\\
&&~~~~~~~~= \langle \kxxnaive \rangle
+ \langle\hat{X} \mathbb{Q} \hat{X} \mathbb{P}H_d\mathbb{P}\rangle + \langle \mathbb{P}H_d \mathbb{P} \hat{X} \mathbb{Q} \hat{X}\rangle, \nn \\
 &&~\tn{where}~\langle \kxxnaive \rangle \equiv \lim_{\alpha \rightarrow 0} \partial_\alpha^2 \langle e^{i \alpha \hat{X}}\mathbb{P} H_d \mathbb{P}e^{-i \alpha  \hat{X}} \rangle. \label{naivegauge}
\eeq
\end{subequations}
We can interpret Eq.~\ref{projgauge} as an ``effective'' gauge transformation associated with the {\it emergent} conserved electronic charge density in the active bands, involving the projected position operator, $\mathbb{P}\hat{X}\mathbb{P}$.  The action of $U_\alpha^{\tn{eff}} \equiv e^{i \alpha \mathbb{P} \hat{X}\mathbb{P}} ~(= e^{i\alpha \sum_{i}\overline{x_i c^\dagger_i c^{\phantom\dagger}_i}})$ on any low-energy state belonging to $\mathbb{H}$ restricts it to the same Hilbert space, where the overline denotes projection to $\mathbb{H}$. The result of Eq.~\ref{projgauge} is thus finite in the limit of $\Delta\rightarrow\infty$ and describes the intrinsic low-energy diamagnetic response associated with only the active bands. On the other hand, $\langle \kxxnaive \rangle$ in Eq.\ref{naivegauge} represents the naive expectation for the diamagnetic response, where the usual gauge transformation $U_{\alpha} = e^{i \alpha  \hat{X}}$ is carried out on the projected Hamiltonian. {\it A priori}, the latter procedure is not even guaranteed to lie solely within $\mathbb{H}$ and clearly differs from the procedure in Eq.~\ref{eq:K_xx_gauge} and \ref{projgauge}. 

It is worth noting that there is a ``partial f-sum rule'' for $T \ll \Delta_t$ that relates $\langle\kxx\rangle$ to the integrated optical spectral weight associated with the interacting isolated bands, i.e.
\begin{equation}
    \int_0^\Lambda d\omega ~\tn{Re} [\sigma^{\tn{eff}}_{\tn{xx}}(\omega)] = \frac{\pi e^2}{2 } \langle \kxx \rangle.
    \label{eq:partial_f_sum}
\end{equation}
Here, the scale $\Lambda$ is chosen with $\widetilde{W}<\Lambda<\Delta$ (Fig.~\ref{bands}b) to obtain the integrated longitudinal optical conductivity for the interacting problem \cite{si}. Thus, the formalism developed in this paper also helps to address the related question of the contribution of the interacting isolated bands to the integrated optical spectral weight at low energies, and with minor modifications can be applied to other problems, e.g. magnetic circular dichroism  \cite{kune_2000,Souza_2008,resta_2020}.

To be explicit, performing the usual gauge transformation generated by $U_{\alpha}$ on the active degrees of freedom (e.g. assuming that there is only one active band) and then projecting back is equivalent to the operation: $c_{\k} \rightarrow \mathbb{P} U_{\alpha}^\dag c_{\k}U_{\alpha} \mathbb{P} = c_{\k+\alpha\vec{e}_x} \langle u_{\k}|u_{\k+\alpha\vec{e}_x}\rangle$, where $\vec{e}_x$ is the unit vector along $x$ direction. This restricted transformation is non-unitary, modifying the measure of the path integral as $|\langle u_\k |u_{\k+\alpha\vec{e}_x}\rangle|^2 \approx 1- \alpha^2 g_{\rm{xx}}(\k)$, where $g_{\rm{xx}}(\k)$ is the quantum metric. This issue of non-unitarity is manifest in some properties of $\langle \kxxnaive \rangle$. First of all, $\langle \kxxnaive \rangle$ is not intrinsic to solely the active degrees of freedom, but instead also depends on the remote bands in the following sense. If we shift the energy levels of the active bands by a constant amount, say $\mathbb{P} \hk \mathbb{P} \rightarrow \mathbb{P} \hk \mathbb{P} + \mathbb{P} \delta$, we have $\langle \kxxnaive \rangle \rightarrow \langle \kxxnaive \rangle - \delta \sum_\k 2 n_\k  g_{\rm{xx}}(\k)$. Moreover, if the active bands are fully filled, the resulting insulator should have a vanishing $\langle\kxx\rangle$; however, $\langle\kxxnaive\rangle$ does not necessarily vanish. To see this dichotomy, note that $e^{i \alpha \mathbb{P} \hat{X} \mathbb{P}}$ is a unitary operator acting on $\mathbb{H}$ and does not change the particle numbers of the active bands. Therefore, if $|\psi\rangle$ is a state with fully filled active bands, $e^{i \alpha \mathbb{P} \hat{X} \mathbb{P}} |\psi\rangle = e^{i \theta_\alpha} |\psi\rangle$, where $\theta_\alpha$ is a phase factor. As a result, $\langle \psi| e^{- i \alpha \mathbb{P} \hat{X} \mathbb{P}} H_d e^{ i \alpha \mathbb{P} \hat{X} \mathbb{P}} |\psi\rangle = \langle \psi | H_d |\psi \rangle$, and $\langle \kxx\rangle =0$. On the other hand, in the same state we obtain $\langle \kxxnaive \rangle = 2 E_0 \sum_{\k,m} g_{\rm{xx}}^{mm} (\k)$, which is in general non-zero. Here $E_0$ is the energy of the insulating state $|\psi\rangle$ and $g_{\rm{xx}}^{mm}(\k)$ is the quantum metric generalized to multiple orbitals, defined in \cite{si}.

{\bf Estimates for microscopic models.-} We can now turn to estimating the different intrinsic contributions from the active bands to $\langle \kxx\rangle$ in order to place an upper bound on $D_s\leq \pi e^2\langle\kxx\rangle$. Specifically, there are three distinct contributions from the active bands to $D_s$ that originate from (i) the bare dispersion, (ii) the interaction-induced dispersion, and (iii) the diamagnetic response due to projected interactions. The first two can be combined together to yield,
\begin{subequations}
\beq
&&\langle\kxx\rangle\bigg|_{\tn{kinetic}} = \sum_\k \frac{\partial^2\widetilde{\ve}_{\k m}}{\partial k_x^2} \langle c^\dagger_{\k m} c^{\phantom\dagger}_{\k m}\rangle,~\tn{where} \label{kxx1}\\
&&\widetilde{\ve}_\k = \ve_{\k m} + \sum_\q V(\q) |\langle u_{\k m}|u_{\k-\q m}\rangle|^2,
\eeq
\end{subequations}
and we have focused on the situation with only one active band (i.e. $m$ is a fixed label) for simplicity. The more general expression appears in \cite{si}. 

The purely interaction induced contribution can be expressed as,
\beq
&&\langle\kxx\rangle\bigg|_{\tn{int}} \label{boundint} \\
&=& \sum_{\k_1,\k_2,\q} V(\q) F(\k_1,\k_2,\q) \langle c_{\k_1m}^\dag c_{\k_2m'}^\dag c^{\phantom\dagger}_{\k_2+\q m'} c^{\phantom\dagger}_{\k_1-\q m} \rangle, \nn
\eeq
where $F(\k_1,\k_2,\q)$ can be expressed as
\beq
F(\k_1,\k_2,\q) =  \left[ \hat{\mathcal{D}}_{\k_1}^x + \hat{\mathcal{D}}_{\k_2}^x\right]^2 \langle u_{\k_1 m} | u_{\k_1-\q m}
\rangle \langle u_{\k_2 m'}| u_{\k_2 +\q m'}\rangle.
\label{F}
\eeq
Here $\hat{\mathcal{D}}_{\k_1}^\mu$ can be viewed as a covariant derivative in $\k-$space acting on the form factor as,
\beq
&&\left[\hat{\mathcal{D}}_{\k}^\mu\right]_{mm',nn'} \langle u_{\k,m'} | u_{\k-\q,n'} \rangle  = \nn\\
&&(\partial_{k_\mu} \delta_{mm'} \delta_{nn'}- i \mathcal{A}_{\k,mm'}^\mu \delta_{nn'} + i\mathcal{A}_{\k-\q,n'n}^\mu \delta_{mm'} ) \langle u_{\k,m'} | u_{\k-\q,n'}\rangle, \nn\\
\eeq
with $\mathcal{A}_{\k,mm'}^\mu = i \langle u_{\k,m} | \partial_{k_\mu} u_{\k,m'}\rangle$ the multi-orbital Berry connection.
We note that given  $\k_1,~\k_2,~\q$ are unrelated to each other, for a generic $V(\q)$, the function $F(...)$ hosts seemingly non-local correlations in momentum space \cite{AA22}. In order to proceed with Eq.~\ref{boundint} further, one option is to evaluate the expectation value in a specific many-body state for a given Hamiltonian. In general, this is difficult as obtaining the true many-body correlations in the state of interest is a challenging affair. An alternative approach is to make  approximations for the four-point expectation values in terms of the band-filling \cite{MR21} and replace each term in the sum by its magnitude to obtain a conservative upper bound on $\langle\kxx\rangle|_{\tn{int}}$ and $T_c$.  

 {\bf Topologically trivial flat-band model.-}  One of the most transparent and straightforward applications of our formalism is to the problem of interacting isolated bands that are  topologically {\it trivial}. Here, we can express $\mathbb{P} H_d \mathbb{P}$ in terms of a basis of exponentially localized Wannier states and obtain the effective electromagnetic response for the interacting theory in the projected Hilbert space. Note that the ``wannierization'' of the underlying Bloch-states is {\it not} an essential element of our framework. To highlight the key insight, we demonstrate the procedure to evaluate $\langle \kxx\rangle$ in the situation with just one active (but possible degenerate) set of bands. Here, we can always choose a gauge where the Berry connection $\mathcal{A}_k \equiv -i \langle u_\k| \partial_{\k} u_\k\rangle = 0$. The effective contribution is then determined by,
\begin{subequations}
\beq
    &&\langle\kxx\rangle = \sum_{\k,\alpha} \partial_{k_x}^2 \ve_{\k,\alpha}~\langle c_{\k,\alpha}^\dag c^{\phantom\dagger}_{\k,\alpha}\rangle \\
    &&~~~~~+ \sum_{\substack{\q,\k_1,\k_2\\\alpha,\beta}}\Xi(\k_1,\k_2,\q) \langle c_{\k_1,\alpha}^\dag c^{\phantom\dagger}_{\k_1-\q,\alpha}  c_{\k_2,\beta}^\dag c^{\phantom\dagger}_{\k_2+\q,\beta} \rangle, \nn\\
  && \Xi(\k_1,\k_2,\q) = V(\q) \left(\partial_{k_{1,x}} + \partial_{k_{2,x}} \right)^2 g_{\alpha \beta}(\k_1,\k_2,\q), \nn\\
\eeq
\end{subequations}
where $g_{\alpha \beta} (\k_1,\k_2,\q) = \langle u_{\k_1,\alpha}| u_{\k_1-\q,\alpha} \rangle \langle u_{\k_2,\beta}| u_{\k_2+\q,\beta} \rangle$ and $\alpha,\beta$ label the spin indices. In terms of the operators $d^{\phantom\dagger}_{i},~d_{i}^\dagger$ corresponding to the Wannier orbitals centered at a site $\vec{r}_i$ spanning the active bands,
\beq
        &&\langle\kxx\rangle = \sum_{i,j} \langle d_{i,\alpha}^\dag d_{j,\alpha}^{\phantom\dagger}\rangle \overbrace{\sum_\k e^{ i \vec{k} \cdot (\vec{r}_i-\vec{r}_j ) } \partial_{k_x}^2 \varepsilon_{\k,\alpha}}^{\mathcal{X}}\nn \\
        &&+ \sum_{\substack{i,j,l,m\\\alpha,\beta}} \langle d_{i,\alpha}^\dag d^{\phantom\dagger}_{j,\alpha}  d_{l,\beta}^\dag d^{\phantom\dagger}_{m,\beta}\rangle \times \nn\\
        &&\underbrace{\sum_{\q,\k_1,\k_2} \Xi(\k_1,\k_2,\q) e^{i \vec{k}_1 \cdot (\vec{r}_i - \vec{r}_j) +i \vec{k}_2 \cdot (\vec{r}_l - \vec{r}_m) -i \vec{q} \cdot (\vec{r}_j - \vec{r}_m) }}_{\mathcal{Y}}.\nn \label{TrivialProj}\\
\eeq
The coefficients $\mathcal{X},~\mathcal{Y}$ can be re-expressed in a more transparent fashion as,
\begin{subequations}
\beq
        &&\mathcal{X} = - (x_i -x_j)^2 \sum_\k e^{ i \vec{k} \cdot (\vec{r}_i-\vec{r}_j ) } \varepsilon_{\k,\alpha} \\
        &&\mathcal{Y} = - (x_i -x_j + x_l - x_m)^2 \times\\
        &&\sum_{\q,\k_1,\k_2} V(\q) e^{i \vec{k}_1 \cdot (\vec{r}_i - \vec{r}_j) +i \vec{k}_2 \cdot (\vec{r}_l - \vec{r}_m) -i \vec{q} \cdot (\vec{r}_j - \vec{r}_m) }  g_{\alpha \beta}(\k_1,\k_2,\q).\nn
\eeq
\end{subequations}
Therefore, for topologically trivial bands, the coefficients $\mathcal{X},~\mathcal{Y}$ can be obtained from the low-energy projected Hamiltonian expressed in terms of Wannier orbitals by carrying out the usual Peierls' substitution, $d_{i,\alpha} \rightarrow d_{i,\alpha} e^{i \vec{A}\cdot \vec{r}_i}$, followed by $\delta^2\heff[A]/\delta A^2$. However, importantly, this procedure is only valid for maximally localized Wannier orbitals \cite{VanderbiltRMP}. 

It is useful to focus on a concrete model, where we can compare our approximate estimate of $D_s^{\tn{upper}} = \pi e^2\langle \kxx\rangle$ with the known $D_s$ evaluated, e.g. using numerically exact quantum Monte-Carlo (QMC) computations. With this in mind, we focus on a two-orbital, spinful time-reversal symmetric model of flat-bands with a tunable quantum metric \cite{heuristicbound}. In the presence of local competing interactions, sign-problem free QMC calculations have revealed a number of intertwined orders \cite{hofmann2022superconductivity}. The Hamiltonian is given by $H=\hk+\hi$,
\begin{subequations}
\beq
\hk &=& -t \sum_{\k} \hat{\mathbf{c}}^{\dagger}_{\k} \left(\tau_x \sin{\alpha_{\k}} + \sigma_z \tau_y \cos{\alpha_{\k}} + \mu \tau_0 \right) \hat{\mathbf{c}}^{\phantom{\dagger}}_{\k},~\tn{where}\nn \\
\alpha_{\k}&=&\zeta [\cos({k_x a})+\cos({k_y a})],\\
\hi &=& - \frac{U}{2} \sum_{\r,l}  \delta \hat{n}_{\r,l}^2
    + V \sum_{\langle \r,\r'\rangle,l} \delta \hat{n}_{\r,l} ~\delta \hat{n}_{\r',l} \,. \label{chiral}
\eeq 
\end{subequations}
Here, $\hat{\mathbf{c}}^{\dagger}_{\k}$ represents the electron creation operators with momentum $\k$, and additional labels --- spin $s=\uparrow,\downarrow$ and orbital $l=1,2$. The Pauli-matrices $\sigma_{j}$ and $\tau_{j}$ act on the spin and orbital indices, respectively. We set the lattice constant, $a=1$ henceforth.
The model is engineered to yield two flat-bands with $\ve_\k=\pm t$ irrespective of the values of $t$ and the  parameter $\zeta$; the latter directly controls the minimal spatial extent associated with the exponentially localized Wannier functions. It is also worth noting that for the present model, the Berry curvature vanishes identically everywhere in the Brillouin-zone, while the Fubini-Study metric is finite and integrates to $\zeta^2/2$. Finally, $\hi$ includes an on-site attraction ($U>0$) and a nearest-neighbor interaction with $\delta \hat{n}_{\r,l} = \sum_s \hat{c}^{\dagger}_{\r,l,s}\hat{c}^{\phantom{\dagger}}_{\r,l,s} - 1$. 


 For this model, we can evaluate $\langle \kxx\rangle$ for the active spin-degenerate flat-bands by choosing the gauge where $\mathcal{A}_k \equiv -i \langle u_\k| \partial_{\k} u_\k\rangle = 0$, and with $V(\q) = U + V [\cos q_x + \cos q_y]$. 
To first gain intuition into what the expansion in Eq.~\ref{TrivialProj} represents, we can fourier-transform these terms back to real-space. 
Specifically, these include contributions to $\langle\kxx\rangle$ from an interaction-mediated hopping term, $\langle\kxx\rangle|_{\tn{hop}}$, and a pair-hopping term, $\langle\kxx\rangle|_{\tn{pair}}$, respectively,
\begin{subequations}
\beq
 \la\kxx\ra|_{\tn{hop}} &\sim& \sum_{i,a_1,a_2,\sigma} (x_{a_1} - x_{a_2})^2\la d_{i,\uparrow}^\dag d^{\phantom\dagger}_{i,\uparrow} d_{i + a_1,\sigma}^\dag d_{i + a_2,\sigma}\ra,\nn\\ \\
   \la\kxx\ra|_{\tn{pair}} &\sim&  \sum_{\langle i,j\rangle} (2 x_i - 2 x_j)^2 d_{i,\uparrow}^\dag d_{i,\downarrow}^\dag d^{\phantom\dagger}_{j,\downarrow} d^{\phantom\dagger}_{j,\uparrow}.
    \label{eq:pair_hop}
\eeq
\end{subequations}
The above exercise also makes it clear that the density-density interactions that are generated within $\heff$ (originating from the $V-$term), do not contribute to $\kxx$ as these do not couple to the vector-potential. 

\begin{figure}
    \centering
    \includegraphics[width = 0.5\textwidth]{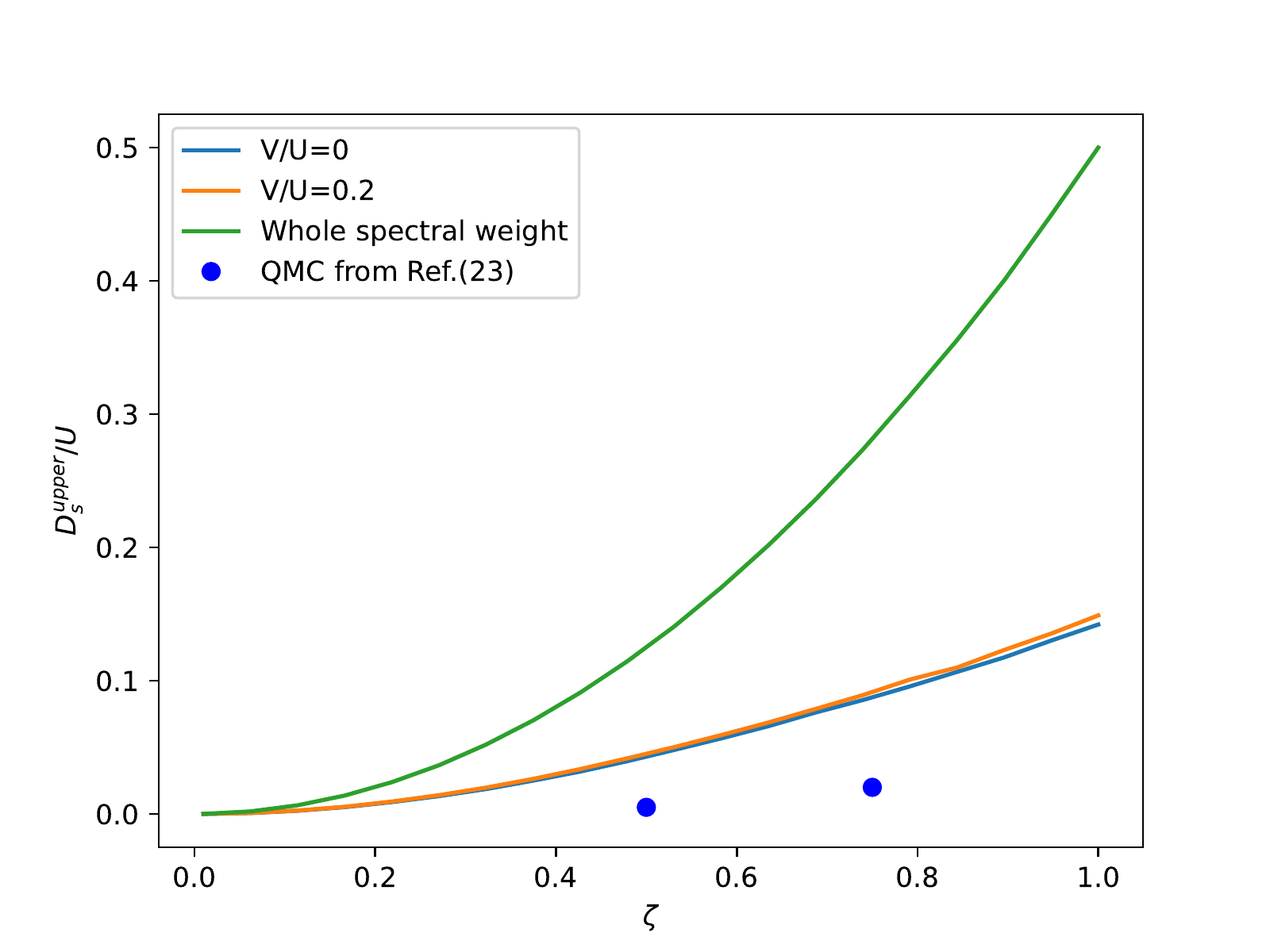}
    \caption{Estimate of $D_s^{\rm{upper}}$ for the interacting flat-band model in Eq.~\ref{chiral} evaluated at $T=0$ and for quarter-filling as a function of $\zeta$. The QMC results for $D_s(0)$ are taken from Ref.~\cite{hofmann2022superconductivity} for $V=0$. The whole spectral weight (green curve) is set by the total bandwidth $t$, and includes contributions from both flat-bands at $\ve_\k=\pm t$. The optical spectral weight for the interaction projected to the lower flat-band, $\ve_\k=-t$, is shown for two different values of $V$ (blue and orange curves). We have taken $t = 4U$ for comparison of the low-energy bound with the bound based on whole spectral weight.}
    \label{fig:trivial-model}
\end{figure}

Finally we turn to an explicit numerical evaluation of $D_s^{\tn{upper}}$ at $T=0$ directly in momentum-space, and a comparison with the exact results obtained using QMC \cite{hofmann2022superconductivity}. Note that for this flat-band model, there is no bound on $T_c/E_F$ with an appropriately defined $E_F$ \cite{heuristicbound}. We plot $D_s^{\tn{upper}}$ w.r.t $\zeta$ in Fig.~\ref{fig:trivial-model}. First, we note that in the limit of $\zeta\rightarrow0$, the sites are disconnected completely such that the optical spectral weight (and relatedly the diamagnetic response) vanishes identically. We can first evaluate the upper bound as determined by the total spectral weight (green line in Fig.~\ref{fig:trivial-model}), which includes both bands at $\ve_\k=\pm t$ and the inter-band matrix-elements. For small $\zeta$, this spectral weight scales as $\sim \zeta^2 t$. Turning to the upper bound as determined by $\hi$ projected to the lower flat-band, the effective $D_s^{\tn{upper}}$ is much closer in magnitude to the result from QMC \cite{hofmann2022superconductivity}. It is interesting to note that for a finite $V$, $D_s^{\tn{upper}}$ is modified, in part due to contributions from a term such as $ \la\kxx\ra|_{\tn{hop}}$. However, in the actual QMC computations, the ground-state is also more susceptible to forming a charge-density wave with increasing $V$ \cite{hofmann2022superconductivity}; the finite result for $D_s^{\tn{upper}}$ guarantees an integrated optical spectral weight via the ``partial f-sum rule'' but does not guarantee an actual superconducting ground-state. 

 We end by noting that recent work \cite{MR21} has carried out a similar analysis on models with local on-site interactions projected to flat-bands, and obtained the associated pair-hopping terms for a ``wannierized'' description.  In general, for an arbitrary gauge choice, or for topological bands where there is an obstruction to constructing exponentially localized Wannier orbitals, one has to use the full gauge-invariant expression for $\kxx$.

{\bf Outlook.-} The central goal of this paper has been to derive the effective electromagnetic response function vis-\`a-vis the phase stiffness (or, the integrated optical spectral weight), in terms of only the low-energy degrees of freedom in the active isolated bands of interest. The formulas derived here can be combined with sophisticated numerical methods to evaluate these quantities accurately. Given the generality of the formalism, it can be applied to a wide range of interacting Hamiltonians, including lattice interacting tight-binding models, and momentum-space models. Similar methods can also be applied to address bounds on the spin-stiffness for interacting ferromagnets in flat-band systems.  A number of recent theoretical works have studied the interplay of berry curvature distribution and quantum geometry in interacting topological flat-bands on the stability of fractional Chern insulators \cite{PJL20,JW21,EB22,TO21}. Finding the fundamental ingredients that guarantee a robust superconducting ground-state remains an interesting problem for the future. We end by noting that we have derived the diamagnetic response from the effective Hamiltonian, $\heff[A]$, in the presence of a constant (probe) vector potential $A$. The more general response functions in the low energy effective theory for a spacetime dependent vector potential have a rich structure and is left for future work.

{\it Acknowledgements.-} DC thanks E. Berg, J. Hofmann and S. Kivelson for related collaborations and for a number of illuminating discussions. DM thanks D. Parker for discussions and feedback on an earlier version of this manuscript. We also thank G. Murthy and M. Randeria for useful discussions. DM is supported by a Bethe/KIC postdoctoral fellowship at Cornell University. DC is supported by a faculty startup grant at Cornell University. DC acknowledges the support provided by the Aspen Center for Physics where this work was completed, which is supported by National Science Foundation grant PHY-1607611.

\bibliographystyle{apsrev4-1_custom}
\bibliography{ref_final}
\clearpage
\renewcommand{\thefigure}{S\arabic{figure}}
\renewcommand{\figurename}{Supplemental Figure}
\setcounter{figure}{0}
\begin{widetext}
\appendix
{\bf \centering Supplementary material for ``Diamagnetic response and phase stiffness for interacting isolated narrow bands"}

\section{Paramagnetic and diamagnetic contributions from projected Hamiltonian}

In this section, we provide additional details on the computation of the electromagnetic response for the effective Hamiltonian, $\heff[A]$. As noted in the main text, the effective current operator can obtained from $\J_\tn{x} = -\delta \heff[A]/\delta A_{\tn{x}}$, after expanding $\heff[A]$ to first order in $A$, 

\beq
        \J_\mu(\q)& =& \mathbb{P} J_\mu (\q) \mathbb{P} + \frac{1}{2}  \sum_{\substack{m,n\in\mathbb{H},\\ \l\notin \mathbb{H}}}  \left(\langle m |H_o |\l\rangle \langle \l| J_\mu(\q) |n\rangle  + \langle m |J_\mu(\q) |\l\rangle \langle \l| H_o |n\rangle\right)\left(\frac{1}{E_m -E_\l} - \frac{1}{E_\l - E_n}\right)\\
       & =& \mathbb{P} J_\mu (\q) \mathbb{P} - \sum_{\substack{m,n\in\mathbb{H},\\ \l\notin \mathbb{H}}} \left(\langle m |H_o |l\rangle \frac{\langle \l| J_\mu(\q) |n\rangle }{E_\l - E_n} + \frac{\langle m |J_\mu(\q) |\l\rangle}{E_\l - E_m} \langle \l| H_o |n\rangle\right) + O(V^2/\Delta).
\label{eq:J_eff_SW}
\eeq
 In the $\q\rightarrow 0$ limit, since the level mixing current can be written as $J_x(\q \rightarrow 0) =  -i \left[\hat{X}, H_d + H_o\right]$ and $|l\rangle$, $|m\rangle$ and $|n\rangle$ are eigenstates of $H_d$, we have $-i \langle l| \left[\hat{X}, H_d + H_o\right] |n\rangle /(E_l - E_n) = i \langle l| \hat{X} |n\rangle  + O(V/\Delta)$ . Applying this trick, Eq.\ref{eq:J_eff_SW} can be simplified to yield the result in the main text.

The effective diamagnetic contribution can be obtained analogously as $\kxx = \frac12 \frac{\delta^2 \heff[A]}{\delta A_\tn{x} \delta A_\tn{x}}$, after expanding $\heff[A]$ to $O(A^2)$,
\begin{equation}
    \begin{split}
      \kxx =& \mathbb{P} K_{\tn{xx}} \mathbb{P} + \sum_{\substack{m,n\in\mathbb{H},\\ \l\notin \mathbb{H}}} \langle m |J_x (\q\rightarrow 0) |\l\rangle \langle \l| J_x(-\q\rightarrow 0) |n\rangle \left(\frac{1}{E_m -E_\l} - \frac{1}{E_\l - E_n}\right)\\
        &+ \frac12 \sum_{\substack{m,n\in\mathbb{H},\\ \l\notin \mathbb{H}}} \left(\langle m |K_{\tn{xx}} |\l\rangle \langle \l| H_o |n\rangle + \langle m |H_o |\l\rangle \langle \l| K_{\tn{xx}} |n\rangle \right)\left(\frac{1}{E_m -E_\l} - \frac{1}{E_\l - E_n}\right).
    \end{split}
    \label{eq:K_xx_SW}
\end{equation}

Let us rewrite the level mixing current $J_x(\q \rightarrow 0) =  -i \left[\hat{X}, H_d\right] -i \left[\hat{X}, H_o\right]$ and $K_{xx} =- \left[\hat{X},\left[\hat{X}, H_d\right]\right] - \left[\hat{X},\left[\hat{X}, H_o\right]\right]$. Since the energy denominators give $O(1/\Delta)$ contributions Eq.\ref{eq:K_xx_SW} and the level mixing current and diamagnetic term give both $O(1)$ and $O(\Delta)$ contributions, we have $O(\Delta)$, $O(1)$ and $O(1/\Delta)$ terms in Eq.\ref{eq:K_xx_SW}. Keeping only the terms that are of $O(\Delta)$ and $O(1)$,
\begin{equation}
\begin{split}
    \kxx =& - \mathbb{P}\left[\hat{X},\left[\hat{X}, H\right]\right]\mathbb{P}\\
    & + \mathbb{P}\left(-2 \hat{X} \mathbb{Q} H_d \mathbb{Q} \hat{X} + H_d \hat{X} \mathbb{Q} \hat{X} + \hat{X} \mathbb{Q} \hat{X} H_d \right)\mathbb{P}\\
    & + 2 \mathbb{P}\left( H_o \hat{X} \mathbb{Q} \hat{X} + \hat{X} \mathbb{Q} \hat{X} H_o -  \hat{X} H_o \hat{X}\right) \mathbb{P}\\
    &+  \mathbb{P}\left(\hat{X} \mathbb{P} \hat{X} H_o+ H_o \hat{X} \mathbb{P} \hat{X} - \hat{X} \mathbb{Q} \hat{X} H_o -H_o \hat{X} \mathbb{Q} \hat{X} \right)\mathbb{P}\\
    =&- \left[\mathbb{P}\hat{X}\mathbb{P}, \left[\mathbb{P}\hat{X}\mathbb{P}, \mathbb{P} H_d \mathbb{P}\right]\right].
\end{split}
\label{eq:K_xx_final}
\end{equation}
 In the above derivation, the term in the first line comes from $\mathbb{P} K_{\tn{xx}} \mathbb{P}$ in Eq.\ref{eq:K_xx_SW} and the rest terms come from the other terms in Eq.\ref{eq:K_xx_SW}.  By expanding $\mathbb{P} K_{\tn{xx}} \mathbb{P}$, we also note that the $O(\Delta)$ piece is $2 \mathbb{P} \hat{X} \mathbb{Q} H_d \mathbb{Q} \hat{X}$, which cancels exactly the first term in the second line in Eq.\ref{eq:K_xx_final}.

\section{Explicit expression for diamagnetic contribution to phase stiffness}

In this section, we express $\kxx$ explicitly in terms of the fields, $c_{\k m},~c^\dagger_{\k m}$, defined in the active bands, where $m$ is the band index. As already emphasized in the main text, the expectation value, $\langle \kxx\rangle$, still depends on the many-body state of interest and is in general difficult to evaluate exactly for a generic model. However, the following exercise will still lead to new insights into the general structure of the theory that controls the diamagnetic contribution. 

First, we consider the effect of the unitary transformation on the field $c_{\k m}$:
\begin{equation}
    \begin{split}
      e^{i\alpha \mathbb{P} \hat{X}\mathbb{P}} c_{\k m} e^{-i\alpha \mathbb{P} \hat{X}\mathbb{P}} = \sum_{\substack{\k'\\m'\in act}}c_{\k' m'} \langle \k,m| e^{-i \alpha  \bar{\hat{x}} }|\k',m'\rangle, ~\tn{where}\\
      \bar{\hat{x}} \equiv \sum_{\substack{\k,\k'\\m,m'\in act}} |\k,m\rangle\langle \k,m| \hat{x}|\k',m'\rangle\langle \k',m'|
    \end{split}
\end{equation}
is the single particle ``projected" position operator. Here ``$act$" is a short form to denote the ``active" bands. We define $|\k,m\rangle \equiv e^{i \k \cdot \vec{x} } |u_{\k,m}\rangle$ as the Bloch wave function. Note that we will be interested in the terms in the above expansion upto $O(\alpha^2)$ and the limit of $\alpha\rightarrow0$.

 It is readily seen that,
\beq
        \langle \k,m | e^{i \alpha  \bar{\hat{x}}} |\k',m'\rangle - \langle \k,m | e^{i \alpha  \hat{x}} |\k',m'\rangle &=& \frac12 \alpha^2 \langle \k,m| \hat{x} \bigg(\sum_{\substack{\k''\\m''\notin act}} |\k'',m''\rangle \langle \k'',m''|\bigg) \hat{x} |\k',m'\rangle + O(\alpha^3)\\
        & =& \frac12 \alpha^2 \delta_{\k,\k'} g_{xx}^{mm'}(\k) + O(\alpha^3),~\tn{where}\\
         g_{\mu\nu}^{mm'}(\k) &=& \left[\langle \partial_\k^\mu u_{\k,m}|\partial_\k^\nu u_{\k,m'}\rangle - \sum_{n\in act} \langle \partial_\k^\mu u_{\k,m}|u_{\k,n}\rangle\langle u_{\k,n}|\partial_\k^\nu u_{\k,m'}\rangle\right],
    \label{eq:U1_diff}
\eeq
with $g_{\mu\nu}^{mm'}(\k)$ the quantum-metric generalized to multiple orbitals. Therefore, we have 
\begin{equation}
    \begin{split}
        e^{i\alpha \mathbb{P} \hat{X}\mathbb{P}} c_{\k m} e^{-i\alpha \mathbb{P} \hat{X}\mathbb{P}} 
        = \sum_{m'\in act }\left\{c_{\k+\alpha\vec{e}_x,m'} \langle u_{\k,m}|u_{\k+\alpha\vec{e}_x,m'}\rangle + \frac12 \alpha^2  g_{xx}^{mm'}(\k) c_{\k,m'}\right\} + O(\alpha^3),
    \end{split}
\end{equation}
where $\vec{e}_x$ is the unit vector along $x$ direction.

We can now obtain the corresponding transformation of the Hamiltonian and the associated diamagnetic response in the main text. For the kinetic energy, $\hk^{\tn{eff}} = \sum_{\k,m\in act}\epsilon_{\k,m} c_{\k,m}^\dag c_{\k,m}$, where $\epsilon_{\k,m}$ includes both the bare dispersion and the interaction induced renormalization,
\beq
    \begin{split}
        e^{i\alpha \mathbb{P} \hat{X}\mathbb{P}} \hk^{\tn{eff}} e^{-i\alpha \mathbb{P} \hat{X}\mathbb{P}} &= \sum_{\substack{\k\\n_1,n_2\in act}} c_{\k,n_1}^\dag c_{\k,n_2} \Big[ \sum_{m\in act} \langle u_{\k,n_1} | u_{\k-\alpha\vec{e}_x,m}\rangle \epsilon_{\k-\alpha\vec{e}_x,m} \langle u_{\k-\alpha\vec{e}_x,m}|u_{\k,n_2}\rangle \\
        &~~~~~~~~~~~~~~~~~~~~~~~~~~~~+\frac12 \alpha^2 g_{xx}^{n_1 n_2}(\k) \left(\epsilon_{\k,n_1} + \epsilon_{\k,n_2}\right) \Big] + O(\alpha^3).
    \end{split}
\eeq

Therefore,
\begin{equation}
    \begin{split}
        \langle\kxx\rangle\bigg|_{\tn{kinetic}} \equiv\partial_\alpha^2 \left(e^{i\alpha \mathbb{P} \hat{X}\mathbb{P}} \hk^{\tn{eff}} e^{-i\alpha \mathbb{P} \hat{X}\mathbb{P}}\right)
        =\sum_{\substack{\k\\n_1,n_2 \in act}} \langle c_{\k,n_1}^\dag c_{\k,n_2}\rangle \left[\langle u_{\k,n_1}| \partial_{k_x}^2 \bar{\epsilon}_\k |u_{\k,n_2}\rangle + g_{xx}^{n_1n_2}(\k) \left(\epsilon_{\k,n_1} + \epsilon_{\k,n_2}\right) \right],
    \end{split}
    \label{eq:k_eff_K}
\end{equation}
 where the operator $\bar{\epsilon}_\k \equiv \sum_{m\in act} \mathbb{P}_{\k,m} \epsilon_{\k,m}$ and $\mathbb{P}_{\k,m} \equiv |u_{\k,m}\rangle \langle u_{\k,m}|$, is a single body projector defined in terms of the Bloch functions. Note that the above quantity does not depend on the bare energy levels $\epsilon_{k,n}$ of the active bands; if we shift $\epsilon_{\k,n}$ by a constant, the two terms in Eq.\ref{eq:k_eff_K} give opposite contributions and cancel each other. When there is only one active band, there is a significant simplification, leading to the familiar results.

Next, we consider the interaction term $\hi^{\tn{eff}}$. To simplify the notation, the repeated indices are summed over and the summation over the band indices only includes the active bands if not specified.
\begin{equation}
    \begin{split}
       &e^{i\alpha \mathbb{P} \hat{X}\mathbb{P}}\hi^{\tn{eff}} e^{-i\alpha \mathbb{P} \hat{X}\mathbb{P}} \\
       &= \sum_{\q,\k_1,\k_2} V(\q) c_{\k_1,\alpha, n_1}^\dag c_{\k_2,\beta, m_1}^\dag c_{\k_2+\q,\beta,m_2} c_{\k_1-\q,\alpha,n_2}\\
        \times \Bigg[&\langle u_{\k_1,n_1} | u_{\k_1 - \alpha,n}\rangle\langle u_{\k_1-\alpha,n}| u_{\k_1-\alpha\vec{e}_x-\q, n}\rangle\langle u_{\k_1-\alpha\vec{e}_x-\q,n}|u_{\k_1-\q,n_2}\rangle \\
        &\times\langle u_{\k_2,m_1} | u_{\k_2 - \alpha,m}\rangle\langle u_{\k_2-\alpha,m}| u_{\k_2-\alpha\vec{e}_x+\q, m}\rangle\langle u_{\k_2-\alpha\vec{e}_x+\q,m}|u_{\k_2+\q,m_2}\rangle \\
        &+ \frac{1}{2}\alpha^2 \Big( g_{xx}^{n_1,n_2}(\k_1) \langle u_{\k_1,n_2}| u_{\k_1-\q, n_2}\rangle + g_{xx}^{n_1,n_2}(\k_1-\q) \langle u_{\k_1,n_1}| u_{\k_1-\q, n_1}\rangle \Big) \langle u_{\k_2,m_1} | u_{\k_2+\q,m_1}\rangle \delta_{m_1,m_2}\\
        &+ \frac{1}{2}\alpha^2 \langle u_{\k_1,n_1} | u_{\k_1-\q,n_1}\rangle \delta_{n_1,n_2} \Big(g_{xx}^{m_1,m_2}(\k_2) \langle u_{\k_2,m_2}| u_{\k_2+\q, m_2}\rangle + g_{xx}^{m_1,m_2}(\k_2+\q) \langle u_{\k_2,m_1}| u_{\k_2+\q, m_1}\rangle\Big) \Bigg]
    \end{split}
\end{equation}
where $\alpha,\beta$ label the spin indices and repeated band indices are summed over active bands. The contribution to $\langle\kxx\rangle|_{\tn{int}}$ is,
\begin{equation}
    \begin{split}
       &\langle\kxx\rangle\bigg|_{\tn{int}}\equiv\partial_\alpha^2 \left(e^{i\alpha \mathbb{P} \hat{X}\mathbb{P}} \hi^{\tn{eff}} e^{-i\alpha \mathbb{P} \hat{X}\mathbb{P}}\right) \\=& \sum_{\q,\k_1,\k_2} V(\q) \langle c_{\k_1,\alpha, n_1}^\dag c_{\k_2,\beta, m_1}^\dag c_{\k_2+\q,\beta,m_2} c_{\k_1-\q,\alpha,n_2}\rangle\\
       &\times\Bigg[2 \langle u_{\k_1,n_1} | \partial_x(\mathbb{P}_{\k_1,n}\mathbb{P}_{\k_1-\q,n})|u_{\k_1-\q,n_2}\rangle \langle u_{\k_2,m_1}|\partial_x(\mathbb{P}_{\k_2,m}\mathbb{P}_{\k_2+\q,m}) | u_{\k_2+\q,m_2}\rangle \\
      &+\langle u_{\k_1,n_1} | \partial_{x}^2(\mathbb{P}_{\k_1,n}\mathbb{P}_{\k_1-\q,n})|u_{\k_1-\q,n_2}\rangle \langle u_{\k_2,m_1} | u_{\k_2+\q,m_2}\rangle \delta_{m_1 m_2}+ \langle u_{\k_1,n_1} | u_{\k_1-\q,n_2}\rangle \delta_{n_1 n_2}\langle u_{\k_2,m_1} | \partial_{x}^2(\mathbb{P}_{\k_2,m}\mathbb{P}_{\k_2+\q,m})|u_{\k_2+\q,m_2}\rangle\\
       &+\Big( g_{xx}^{n_1,n_2}(\k_1) \langle u_{\k_1,n_2}| u_{\k_1-\q, n_2}\rangle + g_{xx}^{n_1,n_2}(\k_1-\q) \langle u_{\k_1,n_1}| u_{\k_1-\q, n_1}\rangle \Big) \langle u_{\k_2,m_1} | u_{\k_2+\q,m_1}\rangle \delta_{m_1,m_2}\\
       &+\langle u_{\k_1,n_1} | u_{\k_1-\q,n_1}\rangle \delta_{n_1,n_2} \Big(g_{xx}^{m_1,m_2}(\k_2) \langle u_{\k_2,m_2}| u_{\k_2+\q, m_2}\rangle + g_{xx}^{m_1,m_2}(\k_2+\q) \langle u_{\k_2,m_1}| u_{\k_2+\q, m_1}\rangle\Big) \Bigg]\\
       &\equiv \langle\kxx\rangle\bigg|_{\tn{int},1} + \langle\kxx\rangle\bigg|_{\tn{int},2},
    \end{split}
\end{equation}
where the terms in the first line in the square bracket is denoted as $\langle\kxx\rangle|_{\tn{int},1}$ and the rest are denoted as $\langle\kxx\rangle|_{\tn{int},2}$.

If there is only one active band (possibly degenerate), we have,
\begin{equation}
    \begin{split}
       \langle\kxx\rangle|_{\tn{int},1} &= \sum_{\q,\k_1,\k_2} V(\q) \langle c_{\k_1,\alpha}^\dag c_{\k_2,\beta}^\dag c_{\k_2+\q,\beta} c_{\k_1-\q,\alpha}\rangle \\
        &\times 2 \left[\partial_{k_{1,x}}\langle u_{\k_1} |u_{\k_1-\q}\rangle +i \langle u_{\k_1} |u_{\k_1-\q}\rangle (\mathcal{A}_{\k_1,x} - \mathcal{A}_{\k_1-\q,x}) \right] \left[\partial_{k_{2,x}}\langle u_{\k_2} |u_{\k_2+\q}\rangle +i \langle u_{\k_2} |u_{\k_2+\q}\rangle (\mathcal{A}_{\k_2,x} - \mathcal{A}_{\k_2+\q,x}) \right], \\
    \end{split}
\end{equation}
\begin{equation}
    \begin{split}
        \langle\kxx\rangle|_{\tn{int},2} &= \sum_{\q,\k_1,\k_2} V(\q) \langle c_{\k_1,\alpha}^\dag c_{\k_2,\beta}^\dag c_{\k_2+\q,\beta} c_{\k_1-\q,\alpha}\rangle\\
        &\times \left\{\left[\partial_{k_{1,x}} + i(\mathcal{A}_{\k_1,x} - \mathcal{A}_{\k_1-\q,x})  \right]^2\langle u_{\k_1}|u_{\k_1-\q}\rangle \langle u_{\k_2}|u_{\k_2+\q}\rangle +  \langle u_{\k_1}|u_{\k_1-\q}\rangle \left[\partial_{k_{2,x}} + i(\mathcal{A}_{\k_2,x} - \mathcal{A}_{\k_2+\q,x})  \right]^2\langle u_{\k_2}|u_{\k_2+\q}\rangle \right\}, \\
    \end{split}
\end{equation}
where $\vec{\mathcal{A}}_{\k} = -i \langle u_\k| \partial_{\k} u_\k\rangle$ is the Berry connection. From here one can read off the expression for $F(\k_1,\k_2,\q)$.

\section{Partial f-sum rule}
In this section, we demonstrate that the following partial f-sum rule holds for the longitudinal conductivity at temperature $T \ll \Delta$,
\begin{equation}
    \int_0^\Lambda d\omega~ \tn{Re} [\sigma_{\tn{xx}}(q_x\rightarrow 0, \omega)] = \frac{\pi e^2}{2 } \langle \kxx \rangle,
\end{equation}
where $\Lambda$ is a cut-off frequency that lies in the gap between the active bands and the upper bands.

In the spectral representation,  the real part of the longitudinal conductivity can be written as,
\begin{equation}
    \begin{split}
       \tn{ Re} [\sigma_{\tn{xx}}(q_x\rightarrow 0, \omega)] = \frac{\pi e^2}{\omega} \sum_{m,n} \frac{e^{-\beta E_n} - e^{-\beta E_m}}{Z}\langle n| j_x(q_x\rightarrow 0) |m\rangle \langle m| j_x(-q_x\rightarrow 0) |n\rangle \delta(\omega - E_m + E_n),
    \end{split}
\end{equation}
where $Z$ is the partition function. If we perform an integral over $\omega$, at temperature $T \ll \Delta$, for $|\omega| \lesssim \Delta$, we only need to consider the states $|n\rangle$ and $|m\rangle$ that belong to the low energy Hilbert space $\mathbb{H}$ and therefore,
\begin{equation}
    \begin{split}
        \int_{-\Lambda}^\Lambda d\omega~ \tn{Re} [\sigma_{\tn{xx}}(q_x\rightarrow 0, \omega)] &\approx \pi e^2 \sum_{m,n\in \mathbb{H}} \langle n| j_x(q_x\rightarrow 0) |m\rangle \langle m| j_x(-q_x\rightarrow 0) |n\rangle  \frac{e^{-\beta E_n} - e^{-\beta E_m}}{Z(E_m - E_n)}\\
        &= - \pi e^2 \sum_{m,n\in \mathbb{H}} \langle n| [\hat{X}, H] |m\rangle \langle m| [\hat{X}, H] |n\rangle \frac{e^{-\beta E_n} - e^{-\beta E_m}}{Z(E_m - E_n)}\\
        &= \pi e^2 \sum_{m,n\in \mathbb{H}} \langle n| \hat{X} |m\rangle (E_m - E_n)\langle m| \hat{X} |n\rangle \frac{e^{-\beta E_n} - e^{-\beta E_m}}{Z} \\
        &= \pi e^2  \sum_{n\in \mathbb{H}} \left\{ 2\langle n| \hat{X} \mathbb{P} H \mathbb{P}\hat{X} |n\rangle -  \langle n| \hat{X} \mathbb{P}\hat{X}\mathbb{P}H |n\rangle  - \langle n| H \mathbb{P}\hat{X} \mathbb{P}\hat{X} |n\rangle  \right\} \frac{e^{-\beta E_n}}{Z} \\
        &= \pi e^2 \langle \kxx \rangle.
    \end{split}
\end{equation}
Note that $\tn{Re} [\sigma_{\tn{xx}}(q_x\rightarrow 0, \omega)]$ is even in $\omega$ and $\langle \kxx \rangle$ is the thermal expectation value of $\kxx$.

\end{widetext}

\end{document}